\documentclass[aps,prl,twocolumn,amsmath,amssymb,nofootinbib,showpacs,superscriptaddress]{revtex4-1}
\usepackage[english]{babel}
\usepackage{latexsym}
\usepackage{graphics}
\usepackage{graphicx}
\usepackage{epsfig}
\usepackage{color}
\usepackage{braket}
\usepackage{lipsum}
\usepackage{siunitx}
\usepackage{comment}
\usepackage{amsmath}
\usepackage{relsize}
\usepackage{CJK}
\usepackage{indentfirst}
\usepackage{float}

\usepackage[colorlinks]{hyperref}
\hypersetup{%
  citecolor=cyan,
  linkcolor=blue,
  urlcolor=cyan
}

\makeatletter
\renewcommand{\fnum@figure}{\textbf{Fig.~\thefigure}}
\makeatother

\DeclareMathOperator{\sinc}{sinc}

\begin{document}

\title{An optical Eratosthenes' sieve for large prime numbers}

\begin{abstract}
We report the first experimental demonstration of prime number sieve via linear optics.   The prime numbers distribution is encoded in the intensity zeros of the far field produced by a spatial light modulator hologram, which comprises a set of diffraction gratings whose periods correspond to all prime numbers below 149. To overcome the limited far field illumination window and the discretization error introduced by the SLM finite spatial resolution, we rely on additional diffraction gratings and sequential recordings of the far field. This strategy allows us to optically sieve all prime numbers below $149^2=22201$.

\end{abstract}

\author{B. Li}
\thanks{These authors contributed equally to this Letter.}
\affiliation{Department of Physics, Hong Kong University of Science and Technology, Clear Water Bay, Kowloon, Hong Kong, China}
\author{G. Maltese}
\thanks{These authors contributed equally to this Letter.}
\affiliation{Department of Physics, University of Oxford, Oxford, OX1 3PU, UK}
\author{J.I. Costa-Filho}
\thanks{These authors contributed equally to this Letter.}
\affiliation{Department of Physics, University of Oxford, Oxford, OX1 3PU, UK}
\author{A.A. Pushkina}
\affiliation{Department of Physics, University of Oxford, Oxford, OX1 3PU, UK}
\author{A.I. Lvovsky}
\email{alex.lvovsky@physics.ox.ac.uk}
\affiliation{Department of Physics, University of Oxford, Oxford, OX1 3PU, UK}
\affiliation{Russian Quantum Center, 100 Novaya St., Skolkovo, Moscow, 143025, Russia}
\affiliation{P. N. Lebedev Physics Institute, Leninskiy prospect 53, Moscow, 119991, Russia}

\maketitle

The study of prime numbers, although millenia old, is still a trending topic in mathematics. On the theoretical side, there are relevant fundamental problems which remain open, such as the Riemann hypothesis and the primes' distribution \cite{ingham1990distribution}. On a more applied note, prime factorization, which is the cornerstone of modern cryptography, motivates a search for more efficent algorithms.

In the past decades, physics emerged as a fruitful venue for progress on prime numbers research. Most efforts were focused on factorization, culminating in the celebrated Shor's algorithm \cite{shor} and the current quest for practical quantum computers. However, the application of analogue physical tools for more fundamental mathematical problems is also being investigated, for example with respect to the Riemann hypothesis \cite{schumayer2011colloquium,bender2017hamiltonian}.

A new and relatively unexplored testbed for the connection between physics and prime numbers is classical optics. On the factorization front, multi-wavelength interferometry \cite{clauser1996factoring} and optoelectronically assisted algorithms \cite{shamir1999factoring} schemes have been proposed, and experimental demonstrations were carried out based on the Talbot effect \cite{pelka2018prime} and polychromatic interference in a multipath interferometer \cite{tamma2011factoring}. On the fundamental side, Berry \cite{berry2012riemann, berry2015riemann} proposed the reading of Riemann function's zeros from radiation patterns using the property of the far field propagation to represent the Fourier transform. 

Inspired by Berry's proposal, Petersen et al. \cite{petersen2019simple} have recently proposed a physical prime numbers sieve based on the superposition of identical waves. The scheme is based on the sieve of Eratosthenes \cite{crandall2006prime}, one of the oldest algorithms for finding prime numbers. Given the sequence of integers $n$ from $2$ to $M$, Eratosthenes' sieve iteratively discards their multiples, such that all remaining integers lower than $M^2$ are necessarily primes. Ref.~\cite{petersen2019simple} proposes to realize the sieving by gratings of equally spaced $n$ point-like sources: in the far field profile, each grating interferes constructively at locations corresponding to the multiples of $n$, and destructively at other integers. The pattern resulting from all these gratings has the prime numbers encoded in its intensity zeros, and locating them permits the realization of the sieve.

\begin{figure}[b]
	\includegraphics[width=1\columnwidth]{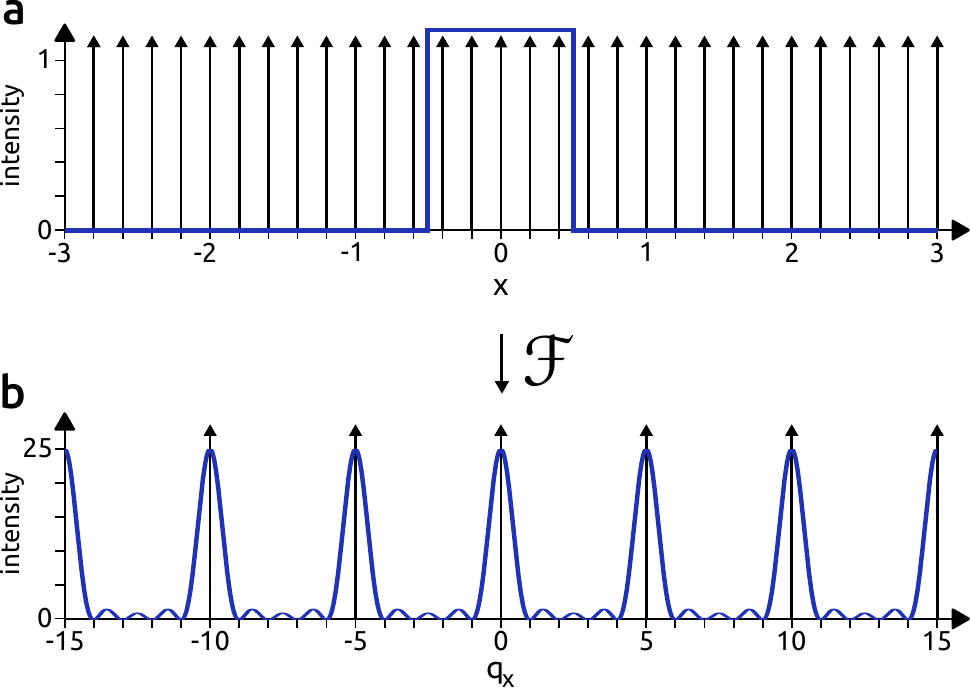}
	\caption{Light intensity profiles at near (a) and far (b) fields, for $n=5$ pinholes per unit length. Black arrows represent an infinite array of pinholes and the corresponding effect on the far field. When this array is restricted to five pinholes [which corresponds to multiplication by the blue envelope in (a)], the far field pattern becomes convolved with a sinc function, but the intensity at all integers except multiples of 5 remains at zero. }
	\label{fig:deltas}
\end{figure}
\begin{figure*}[t]
	\includegraphics[width=2\columnwidth]{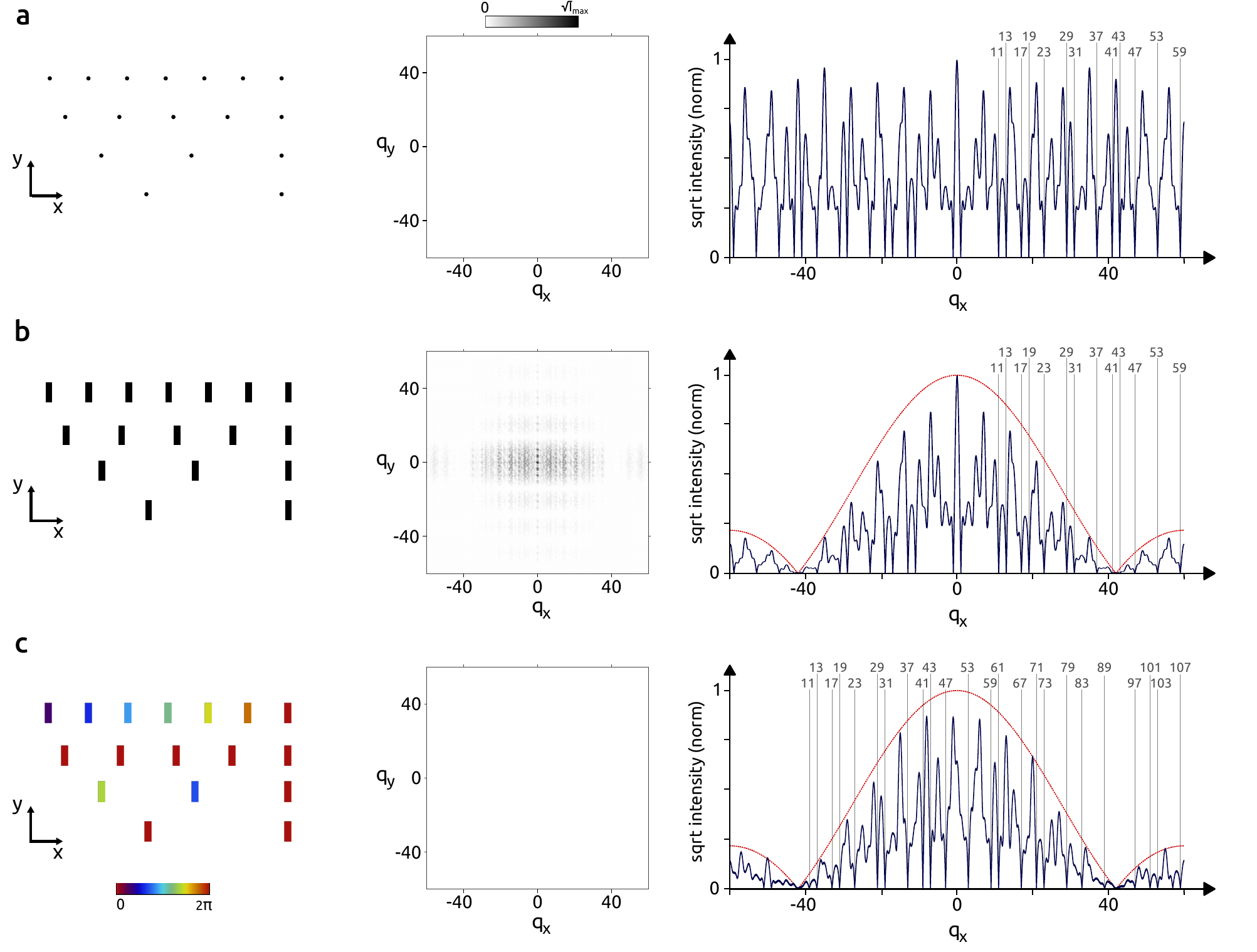}
	\caption{Optical sieve diffraction mask (left) in the case of $\{n\}=\{2,3,5,7\}$. The square root of the corresponding far field intensity profile (centre) and its horizontal marginal (integral over $q_y$, right), which reveals the prime numbers in the intensity zeros, highlighted by vertical grey lines. (a) Sieving mask proposed in \cite{petersen2019simple}, composed of point-like apertures. (b) Sieving mask with finite-size rectangular apertures. (c) Sieving mask superposed with displacement grating, which shifts the primes distribution by $d_x=50$. }
	\label{fig:sketch_apertures}
\end{figure*}

In this paper, we implement and further develop the scheme devised in \cite{petersen2019simple} by means of a phase-only spatial light modulator (SLM), where the apertures are realized by phase masks illuminated by a monochromatic laser. The far field intensity profile is measured by a CMOS camera, and the optical sieve corresponds to intensity minima of that profile. By making use of additional diffraction gratings and sequential camera acquisitions we can extend the sieving range beyond the pixel resolution of the SLM and the camera. The only remaining limitation associated with the SLM pixel resolution is the number of gratings that can be fitted in the phase mask. Capped by this limitation, we demonstrate an optical sieve capable of identifying all prime numbers below $149^2=22201$.


To illustrate the optical sieve proposed in \cite{petersen2019simple}, consider an amplitude grating 
\begin{equation}
g_n(x)=  \sum_{i=0}^{n-1} \delta (x-i/n)
\end{equation}
of $n$ equally spaced point-like sources, aligned along the horizontal direction $x$ and with spatial frequency $1/n$. This can equivalently be seen as an infinite grating of pinholes illuminated by a plane monochromatic wave over a unit length window, as shown in Fig.~\ref{fig:deltas}. Since far field propagation corresponds to taking a Fourier transform
\begin{equation}
\tilde g_n(q_x)=\int_{-\infty}^{\infty} g_n(x)e^{-i 2\pi x q_x}dx,
\end{equation}
where $q_x$ is the horizontal coordinate in the far field plane, this configuration generates a field
\begin{equation}\label{sinc}
\tilde g_n(q_x) =  \left( n \sum_{j=-\infty}^{\infty} \delta (q_x-nj) \right) \ast \sinc(q_x),
\end{equation}
where $\sinc(t) = \sin(\pi t)/ \pi t$, of equally spaced peaks (due to the pinholes) convoluted with a sinc envelope (due to the illumination window). The resulting intensity pattern has zeros at all integers except the multiples of $n$.

Superposing, in the object plane, these gratings $g_n$ for $n \in \{2,3,5,..,M\}$ leads to an interference pattern where, akin to the sieve of Erastothenes, the multiples of the $n$'s are ``discarded" (by becoming intensity peaks) and the remaining zeros correspond to prime numbers. Therefore, via the mapping of the far field zeros, the scheme sieves the prime numbers in the range $M' \leq q_x < M'^2$, where $M'$  is the next prime after $M$. 

The superposition of gratings (which we refer to as the \emph{sieving mask}) can be implemented by displaying each of them at a different vertical position, as shown in Fig.~\ref{fig:sketch_apertures} (left column). In the far field plane, Fig.~\ref{fig:sketch_apertures} (center column), the waves from different gratings interfere, but the intensity at the primes' $q_x$ positions remains zero for all $q_y$. Therefore, a horizontal profile of the far field [Fig.~\ref{fig:sketch_apertures} (right column)], obtained by integration along $q_y$, allows  sieving of the prime numbers as described before.

\begin{figure*}[ht]
	\includegraphics[width=1.98\columnwidth]{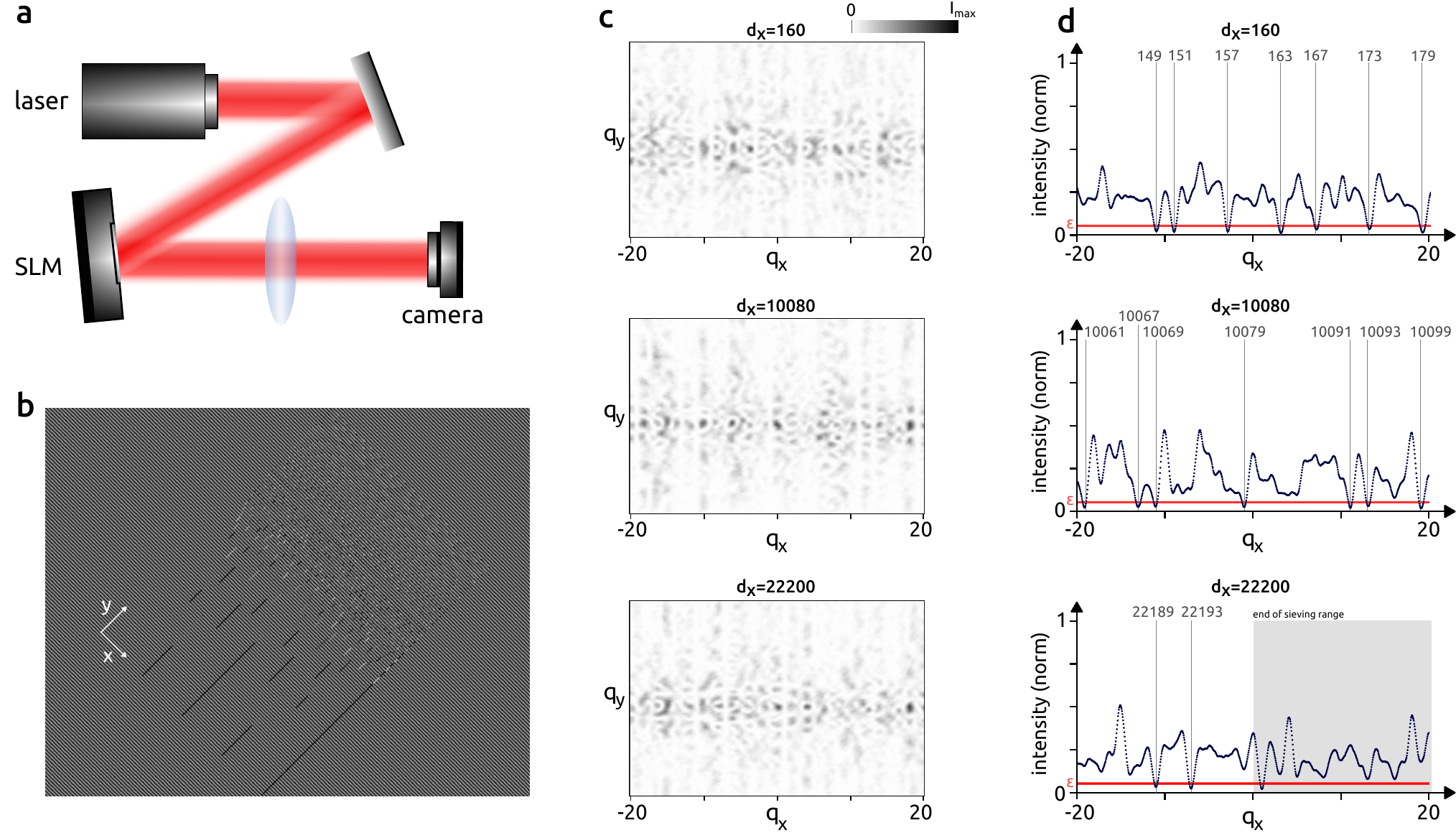}
	\caption{Experimental implementation of the optical prime sieve. a) Scheme of the experiment. b) Phase mask hologram as displayed on the SLM screen. c) Far field intensity images as acquired by the camera for different displacement grating slopes $d_x$. d) Integrals of the acquired intensity images along $q_y$. All minima below the intensity threshold $\epsilon$ (horizontal red line) are classified as prime numbers (vertical grey lines). }
	\label{fig:experimental_results}
\end{figure*}

The above treatment assumes ideal point-like pinholes. A more realistic model would consider finite-size rectangular apertures, as depicted in Fig.~\ref{fig:sketch_apertures}(b), which are mathematically represented by the convolution 
\begin{equation}
E_n(x)=g_n(x)*T_w(x),
\end{equation}
where  $T_w(x)$ is the top-hat function and $w$ is the aperture width. The resulting far field pattern will be modulated by a sinc envelope $\sinc(w q_x)$. This envelope affects the pattern's visibility, which progressively degrades for higher values of $|q_x|$, and also introduces additional zeros, which do not correspond to prime numbers. In practice, it limits the sieve's  working range to the central lobe of the sinc.

This issue is addressed by adding a position-dependent linear function, dubbed here \emph{displacement grating}, to the phase to each aperture [Fig.~\ref{fig:sketch_apertures}(c)]:
\begin{equation}
g_n'(x) = g_n(x) e^{i 2 \pi x d_x}.
\end{equation}
It results in a spatial shift of the intensity zeros by $d_x$ in the $q_x$ coordinate of the far field, 
\begin{equation}\label{displacement}
\tilde g'_n(q_x) = \tilde g_n(q_x-d_x) 
\end{equation}
so that 
\begin{equation}
\tilde E'_n(x)=\tilde g'_n(q_x)\cdot w\sinc(wq_x). 
\end{equation}
Because we keep the phase inside each aperture constant, the modulation envelope  $\sinc(w q_x)$ remains undisplaced. By progressively increasing the slope $d_x$ of the displacement grating we are able to scan different sections of the Eratosthenes' sieve over the central lobe of the sinc envelope, therefore covering the full sieving range.


As illustrated in the scheme of the experimental setup in Fig.~\ref{fig:experimental_results}(a), the optical prime sieve consists of a phase-only liquid crystal on silicon (LCoS) SLM  illuminated almost uniformly by a continuous-wave laser  (Eagleyard EYP-DFB-0785 with $\lambda=785$ nm) beam, along with a camera placed in its far field. 

The spatial profile of the emitted laser beam is filtered by a single mode fiber and magnified by a telescope before illuminating the SLM screen. At the SLM, the laser power is approximately $0.1$mW.The phase-only LCoS-SLM (Hamamatsu X13138-02) has a spatial resolution of $1280 \times 1024$ pixels, $12.5$um pixel pitch, $60$Hz refresh rate and $256$ input signal levels.

A set of optical lenses ($f_1=250$mm, $f_2=75$mm, $f_3=150$mm) produces and magnifies the far field pattern. A monochrome CMOS sensor (uEye UI-3270 CP-M-NO-R2), aligned with the SLM diagonal and anti-diagonal axis, acquires the far field intensity profile. The camera spatial resolution ($2056 \times 1542$ pixels) and the chosen optics allow a proper spatial sampling rate, with approximately $10$ camera pixels corresponding to $\Delta q_x = 1$. The camera's digitiser is set to 12-bit depth and the exposure time to $3.15$ ms, to maximize intensity resolution and avoid saturation.

The sieving mask is implemented by a phase hologram in the SLM screen, as shown in Fig.~\ref{fig:experimental_results}(b). The background outside the ``apertures"  is filled with a blazed grating whose lines are oriented along the $x$ axis. The function of this grating is to diffract the light away from the sieving area, so only the light reflected from the apertures is captured by the camera. The mask is arranged diagonally, in order to prevent the light scattered from both the SLM screen borders and the space between pixels, in the horizontal and vertical directions, from contaminating the image. Additionally, we apply a linear phase displacement  to the apertures, i.e. we multiply $E'_n(x)$ by $e^{i2\pi xd_0}$, which displaces the entire far field pattern by $d_0$, thereby separating it from the SLM screen's specular reflection.

The patterns printed on the SLM extend over $420$ pixels along $x$, accommodating apertures which are $3$ pixels wide. These patterns include gratings corresponding to prime numbers up to $M=139$, allowing a sieving range of $139 < q_x < 149^2=22201$, $149$ being the next prime number after $139$. The displacement gratings' slopes $d_x$ are varied from zero to $22180$ in increments of $40$, covering the whole sieving range.

To improve the contrast, we set the slits' length along $y$ to $\text{round}(200/n)$ pixels, so that each grating $g_n$ contributes equally to the far field intensity pattern. We substitute $g_2$ and $g_3$ by pairs of gratings with $n=4$ and $n=6$, respectively, in order to reduce the space required by the grating with the longest slits. Additional displacement gratings are applied to one element of each pair so the collective action of the pair of elements is equivalent to that of $g_2$ and $g_3$. Further, we extend the total length of $g_4$ by a factor of $1.6$, since it is located close to the SLM screen border and the laser power distribution, which is not perfectly uniform, is slightly lower there.

\begin{figure*}[ht]
	\includegraphics[width=2.0\columnwidth]{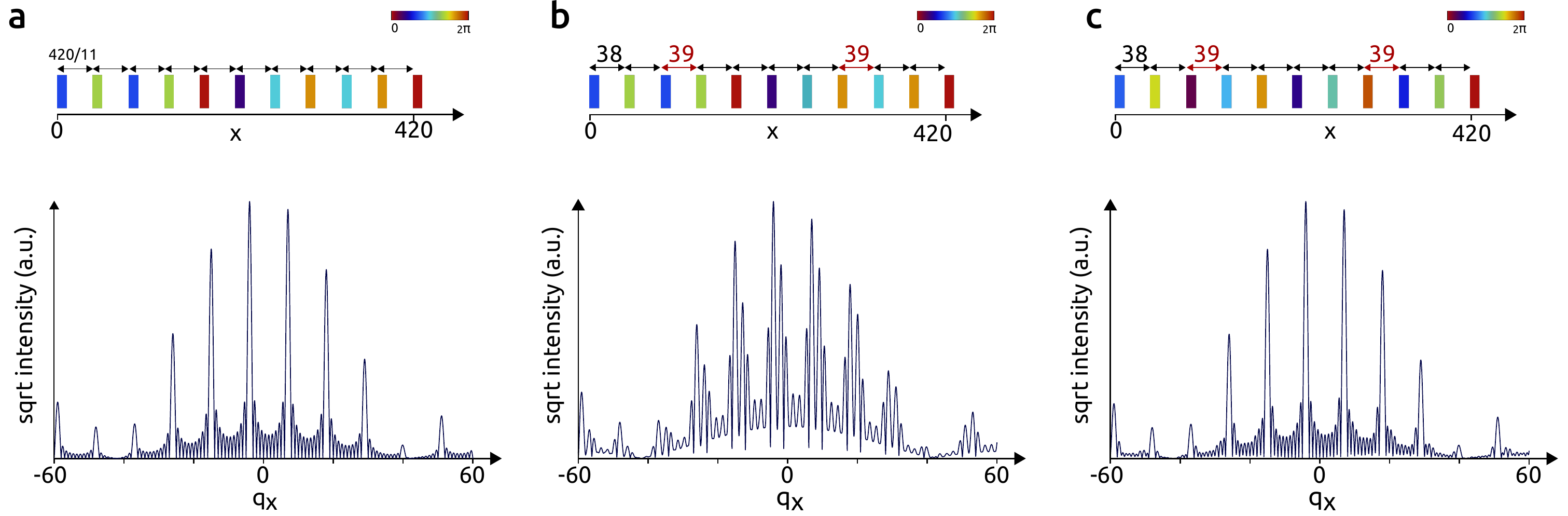}
	\caption{Effect of the SLM's spatial resolution on the produced far field intensity profile for $n=11$ and a displacement grating of slope $d_x=180$ (theoretical model). (a) Evenly spaced apertures (infinite resolution); (b) Discrete pixels with a total pattern width of $W=420$ pixels: the apertures are spaced by either $38$ or $39$ pixels. The integers not multiples of 11 are no longer zeros in the far field intensity profile. (c) $W=420$ with a displacement grating of slope $d^{(11)}_x=180 \mod 11 =4$: the desired structure of the far field is closely restored.
	}
	\label{fig:spatial_resolution_error}
\end{figure*}

A key requirement on the optical prime sieve is the even spacing of its apertures, combined with their precise positioning. This is difficult to achieve in practice given a large number of gratings that need to be printed, as well as the finite SLM pixel resolution. If the apertures' spacing is rounded to the closest integer, the far field pattern becomes severely distorted and can no longer be used for sieving [Fig.~\ref{fig:spatial_resolution_error}(a,b)]. 

This distortion is significant when the function $\tilde g_n(\cdot)$ has a large argument, i.e., according to Eq.~(\ref{displacement}), for large displacement grating slopes $d_x$. We address this issue by recalling that each function $\tilde g_n(\cdot)$ is periodic with zeros at multiples of $n$. Hence displacing it by $d_x$ is equivalent to displacing it by the remainder of the division of $d_x$ by $n$. Therefore the  slope of each displacement grating can be modified according to $d^{(n)}_x =d_x \mod n$
without any effect on the sieve, albeit at the cost of a different displacement slope for each line $g_n$. 

The effect of this correction is illustrated in Fig.~\ref{fig:spatial_resolution_error}(c). This compensation allows us to keep the argument of $\tilde g_n(\cdot)$ low for all $g_n$, thereby achieving correct sieving. 

The aberrations of the setup, due to both the lenses along the optical path and the intrinsic SLM curvature, distort the wavefront of the diffracted beam, reducing its spatial coherence and thus the interference visibility. This is corrected by adding a global phase profile to the SLM mask, which is measured via the calibration method proposed in \cite{vcivzmar2010situ}.

Fig.~\ref{fig:experimental_results}(c) shows the recorded far field intensity image for the SLM hologram of Fig.~\ref{fig:experimental_results}(b). To maximize the contrast between primes zeros and composites minima, each recorded frame  is integrated over $600$ pixels along $q_y$, generating an average intensity plot $I(q_x)$ [Fig.~\ref{fig:experimental_results}(d)]. During this operation, we also subtract the background noise, recorded after applying to the SLM a hologram without the apertures. To cover the whole sieving range, we need to acquire $22000/40=550$ frames. 
When the intensity value $I(q_x)$ is lower than the threshold $\epsilon$, $q_x$ is labeled as a prime.

\begin{figure}[h!]
	\includegraphics[width=0.4\textwidth]{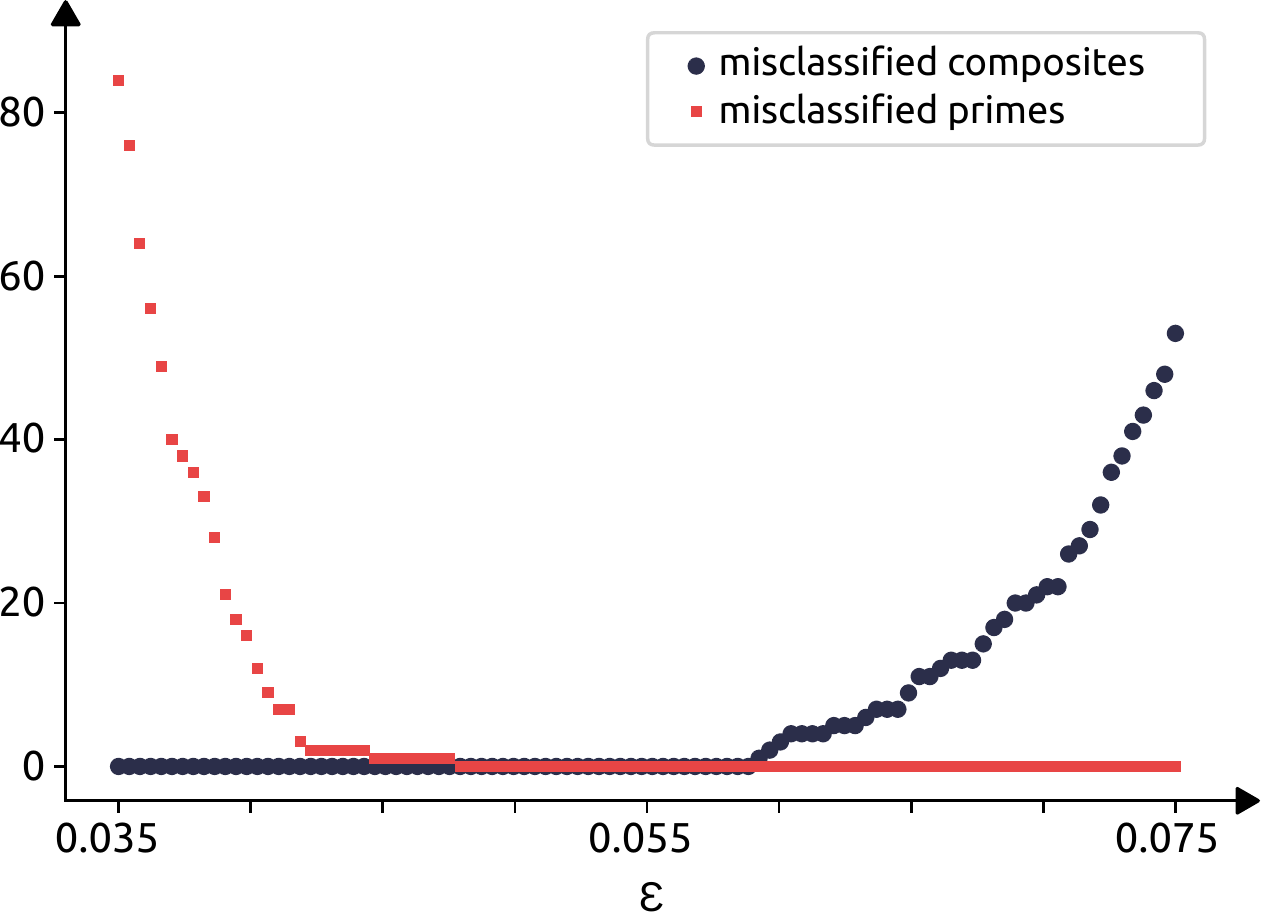}
	\caption{
		Number of misclassified primes (red squares) and composites (blue dots) within the measured sieving range $139 < q_x <22201$ as a function of the intensity threshold $\epsilon$. 
		By choosing $0.048 < \epsilon < 0.059$, the optical sieve correctly sieves all primes and discards composites.}
	\label{fig:epsilon_range}
\end{figure}

We find that an intensity threshold that is in between $4.8\%$ and $5.9\%$ of the global maximum allows us to correctly sieve all the primes ($2436$ numbers) and discard all the composites ($19626$ numbers) within the sieving range (Fig.~\ref{fig:epsilon_range}). This range quantifies the robustness of our scheme.

This work is the first experimental realization of an optical prime number sieve. By generating high resolution and on-demand phase masks via an SLM, we prove the effectiveness of the sieve by correctly classifying integers up to 22201 as either primes or composites.

The scalability of the sieving range, dictated by the number of apertures that can be fitted in a phase mask, is ultimately limited by the SLM spatial resolution. It is also limited by the SLM refresh rate, since a sequence of images has to be displayed in order to cover the whole sieving range. As SLM technology improves with respect to this two features, we expect better optical sieves to be implemented. The principally achievable range of the sieve can be estimated as the square of the SLM pixel resolution. 

Our method can be straightforwardly adapted to other kinds of sieves, such as Gaussian or twin primes sieves \cite{petersen2019simple}. More generally, it demonstrates the potential of free-space linear optics in tasks that transcend imaging or communication, such as computation and solving mathematical problems. We can leverage the spatial coherence and superposition properties of light to achieve parallelism in computing. In the present case,  all prime numbers within a certain range are found by means of a single optical operation rather than multiple multiplications and divisions that a digital electronic computer would require. 





A.L.'s research is partially supported by Russian Science Foundation (19-71-10092).

\bibliographystyle{apsrev4-1}
\bibliography{biblio}

\end{document}